\documentclass[orivec]{llncs}

\usepackage[a4paper,margin=1in,footskip=0.25in]{geometry}

\usepackage{array} 
\usepackage{graphicx}
\usepackage{graphicx}
\usepackage[dvipsnames]{xcolor}
\DeclareRobustCommand{\legendsquare}[1]{%
  \textcolor{#1}{\rule{1ex}{1ex}}%
}
\usepackage{MnSymbol}

\usepackage{stackengine}

\usepackage{subfig}
\usepackage{multirow}
\usepackage[labelfont=bf]{caption}

\usepackage{tikz,xcolor,hyperref}

\definecolor{lime}{HTML}{A6CE39}
\DeclareRobustCommand{\orcidicon}{%
	\begin{tikzpicture}
	\draw[lime, fill=lime] (0,0) 
	circle [radius=0.16] 
	node[white] {{\fontfamily{qag}\selectfont \tiny ID}};
	\draw[white, fill=white] (-0.0625,0.095) 
	circle [radius=0.007];
	\end{tikzpicture}
	\hspace{-2mm}
}

\foreach \x in {A, ..., Z}{%
	\expandafter\xdef\csname orcid\x\endcsname{\noexpand\href{https://orcid.org/\csname orcidauthor\x\endcsname}{\noexpand\orcidicon}}
}

\begin{document}
\title{W-Net: Dense Semantic Segmentation of Subcutaneous Tissue in Ultrasound Images by Expanding U-Net to Incorporate Ultrasound RF Waveform Data}
%

\titlerunning{W-Net:  Dense Semantic Segmentation of Subcutaneous Tissue}
%





\author{Gautam Rajendrakumar Gare\inst{1} \orcidA{} \and
Jiayuan Li \inst{1} \and
Rohan Joshi \inst{1} \and
Mrunal Prashant Vaze \inst{1} \and
Rishikesh Magar \inst{1} \and
Michael Yousefpour \inst{1,2} \and
Ricardo Luis Rodriguez \inst{3} \and
John Micheal Galeotti\inst{1} \orcidB{}}

\authorrunning{G. Gare et al.}
%
\institute{Carnegie Mellon University, Pittsburgh PA 15213, USA 
\email{\{ggare,jiayuanl,rohanj,mvaze,rmagar,myousefp,jgaleotti\}@andrew.cmu.edu}\\ \and
University of Pittsburgh Medical Center, Pittsburgh PA 15260, USA \\
\email{yousefpourm@upmc.edu}\\ 
 \and
Cosmetic Surgery Facility LLC, Baltimore, MD 21093, USA\\ 
\email{dr.rodriguez@me.com}
}

\maketitle
%


\begin{abstract}
We present W-Net, a novel Convolution Neural Network (CNN) framework that employs raw ultrasound waveforms from each A-scan, typically referred to as ultrasound Radio Frequency (RF) data, in addition to the gray ultrasound image to semantically segment and label tissues.  Unlike prior work, we seek to label every pixel in the image, without the use of a background class.  To the best of our knowledge, this is also the first deep-learning or CNN approach for segmentation that analyses ultrasound raw RF data along with the gray image. International patent(s) pending [PCT/US20/37519].
We chose subcutaneous tissue (SubQ) segmentation as our initial clinical goal since it has diverse intermixed tissues, is challenging to segment, and is an underrepresented research area.  SubQ potential applications include plastic surgery, adipose stem-cell harvesting, lymphatic monitoring, and possibly detection/treatment of certain types of tumors.  A custom dataset consisting of hand-labeled images by an expert clinician and trainees are used for the experimentation, currently labeled into the following categories: skin, fat, fat fascia/stroma, muscle and muscle fascia. We compared our results with U-Net and Attention U-Net. Our novel \emph{W-Net}'s RF-Waveform input and architecture increased mIoU accuracy (averaged across all tissue classes) by 4.5\% and 4.9\% compared to regular U-Net and Attention U-Net, respectively. We present analysis as to why the Muscle fascia and Fat fascia/stroma are the most difficult tissues to label. Muscle fascia in particular, the most difficult anatomic class to recognize for both humans and AI algorithms, saw mIoU improvements of 13\% and 16\% from our W-Net vs U-Net and Attention U-Net respectively.

\end{abstract}




\vspace{1cm}
\section{Introduction}
Since its first application in medicine in 1956, ultrasound (US) has become increasingly popular, surpassing other medical imaging methods to become one of the most frequently utilized medical imaging modalities. There are no known side effects of Ultrasound imaging, and it is generally less expensive compared to many other diagnostic imaging modalities such as CT or MRI scans. Consequently, ultrasound implementation for diagnosis, interventions, and therapy has increased substantially \cite{noble:hal-00338658}. In recent years, the quality of data gathered from ultrasound systems has undergone remarkable refinement \cite{noble:hal-00338658}. The increased image quality has enabled improved computer-vision ultrasound algorithms, including learning-based methods such as current deep-network approaches.

In spite of the increased image quality, it can still be challenging for experts (with extensive anatomic knowledge) to draw precise boundaries between tissue interfaces in ultrasound images, especially when adjacent tissues have similar acousto-mechanical properties. This has motivated researchers to come up with techniques to assist with tissue typing and delineation. Most techniques extract features from the grey ultrasound image, that are then used for classification using SVM, random forests, or as part of convolutional neural networks (CNN). Most of the recent algorithms (especially CNN based approaches) differentiate tissues based on visible boundaries \cite{gray_image_based_seg1} \cite{gray_image_based_seg2}. The performance of these algorithms is thus directly dependent on the quality of the ultrasound image.  One prior approach towards differentiating tissue classes in ultrasound is to also utilize the raw RF data from the ultrasound machine \cite{gorce} \cite{rf_based_seg}. The RF data can be analyzed to determine the dominant frequencies reflected/scattered from each region of the image, allowing algorithms to differentiate tissues based on their acoustic frequency signatures rather than just visible boundaries.

The spectral analysis of ultrasound RF waveforms to perform tissue classification has been extensively investigated in the past, primarily using auto-regression for feature generation with subsequent classification  using techniques such as SVM, random forests, and (not deep, not convolutional) neural networks \cite{gorce}. Recently, CNN based approaches that use RF data have emerged, but for other applications such as beamforming and elastography \cite{8432500}.  Joel Akeret et al. \cite{radio} used U-net to mitigate RF interference on 2D radio time-ordered data, but the data was acquired from a radio telescope instead of an ultrasound machine. 
To the best of our knowledge, our work is the \emph{first deep-learning and/or CNN technique which makes use of the RF data for ultrasound segmentation, the first to combine RF data with the grey ultrasound image for the same, and possibly the first to combine waveform and imaging data for any CNN/deep-learning image segmentation or labeling tasks.} (Patents are pending.)

CNN segmentation algorithms for ultrasound are gaining popularity, with better detection of anatomic boundaries and better characterization of tissue regions.  The U-Net architecture developed by O. Ronneberger et al. \cite{unet} was a major milestone in biomedical image segmentation. It provided the first effective technique and now acts as the baseline architecture for current CNN models. More recently, attention mechanisms have been widely used in  machine translation \cite{Bahdanau2014NeuralMT}\cite{attention}, image captioning \cite{DBLP:journals/corr/AndersonHBTJGZ17} and other tasks using the encoder-decoder model. Ozan Oktay et al. \cite{AUnet} extended the U-Net architecture by applying the attention mechanism to the decoding branch on the U-Net using a  self-attention gating module which significantly improved the segmentation accuracy. Attention mechanisms are regarded as an effective way to focus on important features without additional supervision. Our present approach extends the U-Net architecture by adding an RF encoding branch along with the grey image encoding branch.  We  compare our approach with the traditional U-Net and attention U-Net architecture.  (Attention-W-Net is not yet implemented.)
%
%

Unlike prior work \cite{Mishra2018SegmentationOV}\cite{UltrasoundSeg6a}\cite{7950607}, we seek to label every pixel in the image, without the use of a background class.  Such \emph{dense semantic segmentation} paves the way to a more complete understanding of the image. While it has found vast application in autonomous driving, dense semantic segmentation has not yet seen widespread application in the medical field. Medical diagnoses and procedures can potentially benefit from such an exhaustive understanding of the anatomical structure, potentially expanding applications such as detecting cancerous cells, malformations, etc. We hypothesize (but do not evaluate in this work) that labeling every pixel into appropriate categories instead of treating every ``non-interested" pixel as background can help eliminate false positives, which is a common issue encountered in CNN based segmentation techniques \cite{DBLP:journals/corr/abs-1801-05173}.

We chose subcutaneous tissue (SubQ) segmentation as our initial clinical goal. The SubQ space has not yet received substantial interest in medical imaging, and yet it contains many critical structures such as vessels, nerves, and much of the lymphatic system. It presents potential applications in plastic surgery, adipose stem-cell harvesting, lymphatics, and possibly certain types of cancer. Furthermore, since most ultrasound is done through the skin, SubQ tissues are a subset of those tissues present in most other, deeper ultrasound images, and as such provide a minimal on-ramp to achieving dense semantic segmentation now, and then extending it to deeper tissue later.  So, the clinical goal of this study is to delineate the boundaries of structures in the SubQ space and label/characterize those tissue regions. We currently label into following categories: skin (epidermis/dermis combined), fat, fat fascia/stroma, muscle, and muscle fascia. In the future, our approach could also be applied to the delineation of additional classes of tissues and tools, such as vessels, ligaments, lymphatics, and needles, which are of vital interest in carrying out medical procedures.

\subsubsection{Contributions}
Our main technical contribution is (1) a novel RF-waveform encoding branch that can be incorporated into existing architectures such as U-Net to improve segmentation accuracy. Other key contributions are:  (2)	Unlike prior work, we seek to label every pixel in the image, without the use of a background class. (3) To the best of our knowledge, this is the first segmentation task targeting the SubQ region. (4) We demonstrate the benefits of employing RF waveform data along with grey images in segmentation tasks using CNNs. (5)	We present a novel data padding technique for RF waveform data.


\section{Methodology}

\subsubsection{Problem Statement}
Given an ultrasound grey image $I_g$ and RF image $I_{RF}$ containing RF waveform data in each column, the task is to find a function $F \colon [ \, I_g, I_{RF} ] \, \to L$ that maps all pixels in $I_g$ to tissue-type labels $L \in \{1, 2, 3, 4, 5\}$. Because tissues of various densities have distinct frequency responses when scattering ultrasound, the segmentation model should learn underlying mappings between tissue type, acoustic frequency response, and pixel value.  The subcutaneous interfaces presently being segmented are: (1) Skin [Epidermis/Dermis], (2) Fat, (3) Fat fascia/stroma, (4) Muscle and (5) Muscle fascia.

\subsection{Architecture}

 



\begin{figure}[!ht]
\def\big{\includegraphics[height=4.5in, width=\textwidth]{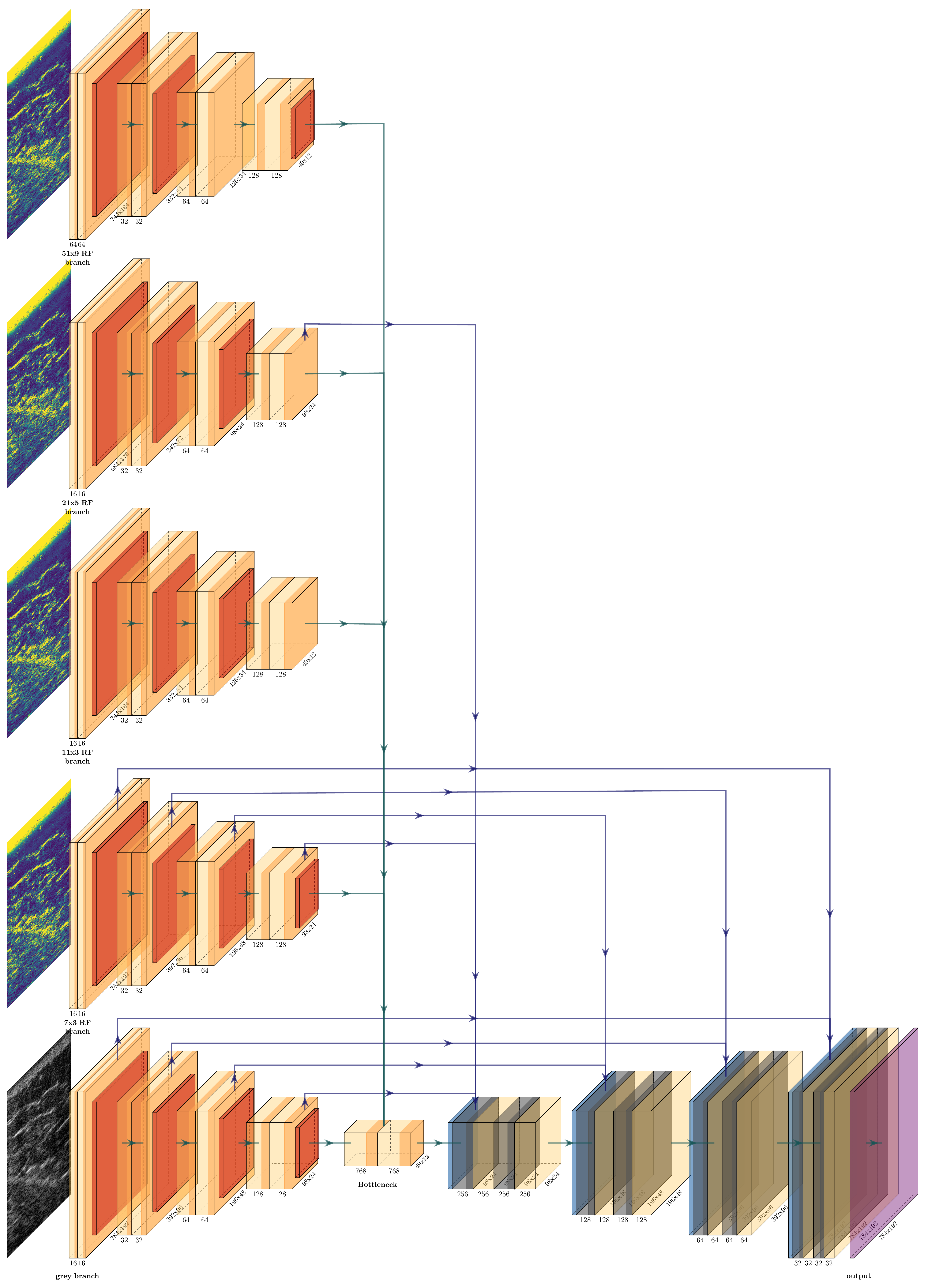}}
\def\little{\includegraphics[height=2.5in, width=0.45\textwidth]{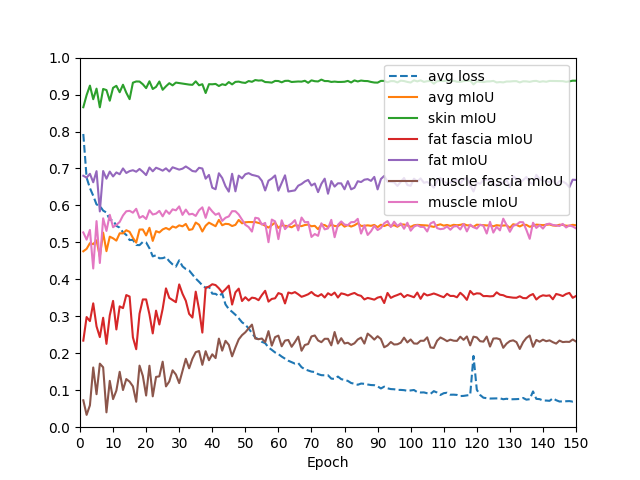}}
\def\stackalignment{r}
\topinset{\little}{\big}{7pt}{7pt}
\caption{ \small W-Net model architecture. 
\legendsquare{Dandelion} conv, batchnorm, ReLU 
\legendsquare{BrickRed} max-pool 
\legendsquare{MidnightBlue} up-conv
\legendsquare{Apricot} conv, batchnorm, ReLU 
\legendsquare{DarkOrchid} conv, softmax
\textbf{Inset:}  Loss and mIoU accuracy vs Epoch.} 
\label{fig:Wnet_model}
\end{figure}

The W-Net architecture is illustrated in Fig. \ref{fig:Wnet_model}. The grey image encoding branch and the decoding branch of W-Net is similar to the U-Net architecture, with the addition of the RF encoding branch(es) described below.

\subsubsection{RF Encoding branch}
RF encoding is implemented using multiple parallel input branches, each with a different kernel size corresponding to different wavelengths. This architecture implicitly leads to binning the RF waveform analysis into different frequency bands corresponding to the wavelength support of each branch, which we hypothesize can aid in classification/segmentation. Tall-thin rectangular kernels were used on the RF data branches so as to better correlate with the RF waveforms of individual A-scans, with minimal horizontal support (e.g. for horizontal de-noising blur).  Each RF branch consists of 4 convolution blocks wherein each block is a repeated application of two convolution layers, each followed by batch normalization and ReLU activation layers, followed by a max-pooling layer, similar to the U-Net architecture \cite{unet}. All of the encoding branches come together at the “bottom” of the U, where the encoding and decoding branches meet.  The outputs from all the RF encoding branches are concatenated along with the output of the grey image encoding branch, which feeds into a single unified decoding branch. Just as U-net uses skip connections between the image encoding and decoding branches, we add additional skip connections from the RF encoding branches into the decoder, implemented at each scale for which the dimensions agree. This can be viewed as a late fusion strategy as often used with multi-modal inputs \cite{UNet_latefusion1} \cite{UNet_latefusion2}. In our case, due to the an-isotropic mismatch between the RF waveform and grey image data, we developed multiple unique downsampling branches for RF and grey image encoding branch by choosing kernel sizes so as to ultimately achieve late fusion at the bottom layer of the U-Net. 

We pre-configured our network to perform spectral analysis of the RF waveform by initializing the weights of the first few layers of the RF encoding branches with various vertically oriented Gabor kernels, encouraging the model to learn  appropriate Gabor kernels to better bin the RF signal into various frequency bands.  Our initial Gabor filters included spatial frequencies in the range $[0.1, 0.85]$ with variance $\sigma_x \in [3, 5, 10, 25]$ and $\sigma_y \in [1,2,4]$. These Gabor filters have frequency separation of 3-8MHz to be  within the range of standard clinical practice, especially for portable point-of-care ultrasound (POCUS).

The first 2 convolution blocks of each RF encoding branch consists of kernels designed to hold a specific size of Gabor filter, of sizes 7x3, 11x3, 21x5, and 51x9 (one per branch). 
The 11x3 Gabor filter was embedded into a convolution kernel of size 21x5 (going out to two standard deviations instead of one).  The 4 kernel sizes  were chosen to allow skip connections into the decoding branch (matching the output size of the grey image encoding branch).
RF encoding branches with 11x3, 21x5, and 51x9 kernels have no max-pooling in Conv block 4, 4, and 3 respectively to compensate for their loosing more input-image boundary pixels.



\subsection{Data}

The proposed method was validated on a custom dataset consisting of ultrasound scans from 2 de-identified patients under different scan settings, obtained using a Clarius L7 portable ultrasound machine, which is a 192-element, crystal-linear array scanner. The dataset consisted of 141 images of dimension 784x192, wherein 69 images are of patient A and remaining 72 images are of patient B. The RF images were in .mat format and had a range of -15K to 15K for patient A and -30K to 30K for patient B. Our small dataset required augmentation, but standard crop and perspective warp techniques could not be applied as it would distort the RF waveform data (by introducing discontinuities in the waveform) which would adversely impact spectral analysis. So, we augmented the data only via left-to-right flipping and scaling the RF and grey image pixels by various scales [0.8, 1.1]. This resulted in a 6 fold increase in the dataset size leading to a total of 846 augmented images, which we divided into 600 training, 96 validation, and 150 testing images.



\subsubsection{Feature Engineering}
The RF images of the two patients were of different dimensions (784x192 and 592x192). Since the CNN architecture is limited to fixed-size input, we zero-padded the smaller labeled and grey images at the bottom to match the input size. To minimize the introduction of phase artifacts when padding the RF images, shorter RF images were mirrored and reflected at the deepest zero crossing of each A-scan to avoid waveform discontinuities, to fill in padding values. Our training and testing error metrics treated the padded region as a special-purpose background in the segmentation task and excluded the region from the loss function while training the neural network.


\subsubsection{Implementation}
The network is implemented with PyTorch and trained with an Adam optimizer to optimize over Cross-entropy loss. A batch size of 4 is used for Gradient update computation. Batch normalization is used.  Due to the small size of our augmented training dataset, which allowed fewer learnable parameters, the number of kernels in each layer for all the models was reduced by a factor of 4 compared to traditional U-Net.

\subsubsection{Metrics}
Mean Intersection over Union (mIoU) and pixel-wise accuracy were the primary metrics used. We calculated mIoU per segmentation category and mean mIoU across all segmentation categories. Since some of the test images (8) did not contain all the tissue classes, we stacked all the test images horizontally and calculated the mean mIoU values reported in Tables \ref{tab:seg_results} \& \ref{tab:dataset_size_results}. 

\section{Experiment-1: Tissue Segmentation}
We compare W-Net's performance against Attention U-Net (AU-Net) \cite{AUnet} and the traditional U-Net \cite{unet} architectures. For the U-Net and AU-Net, we analyze with two combinations of inputs, one with single-channel grey image, the other with a second channel containing RF data.
Table \ref{tab:seg_results} depicts the mean and standard deviation of segmentation pixel-wise and mIoU scores obtained by evaluating on \emph{seven independently trained models}. The scores were calculated on 25 test images augmented with horizontal flipping.
The segmentation results are shown in Fig. \ref{fig:seg_results}. Our W-Net better delineates boundaries between fat and muscle as compared with U-Net and AU-Net.
%
%
%
%
%

\newlength{\ultrasoundwidth}
\setlength{\ultrasoundwidth}{0.75 in}
\newlength{\ultrasoundheightA}
\setlength{\ultrasoundheightA}{0.553 in}
\newlength{\ultrasoundheightB}
\setlength{\ultrasoundheightB}{0.419 in}

\begin{figure}[!ht]
\centering
\newcolumntype{C}{>{\centering\arraybackslash}m{\ultrasoundwidth}<{}}
\begin{tabular}{m{0.3cm}CCCCCCCC}
 &
grey &
RF &
label &
U-Net &
U-Net \newline \scriptsize with RF &
AU-Net &
AU-Net \newline \scriptsize with RF &
W-Net \newline \scriptsize uses RF\\
\subfloat{\rotatebox[origin=lB]{90}{ \scriptsize patient-A}} &
\subfloat{\includegraphics[height = \ultrasoundheightA, width = \ultrasoundwidth]{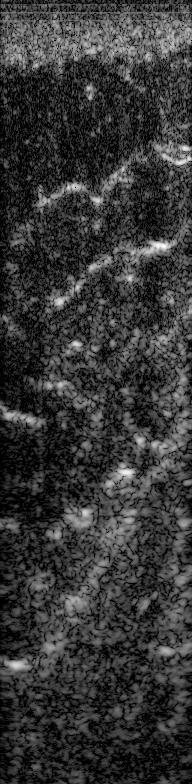}} &
\subfloat{\includegraphics[height = \ultrasoundheightA, width = \ultrasoundwidth]{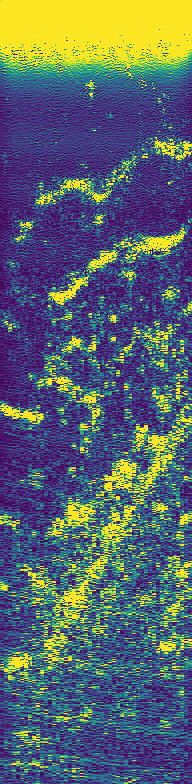}} &
\subfloat{\includegraphics[height = \ultrasoundheightA, width = \ultrasoundwidth]{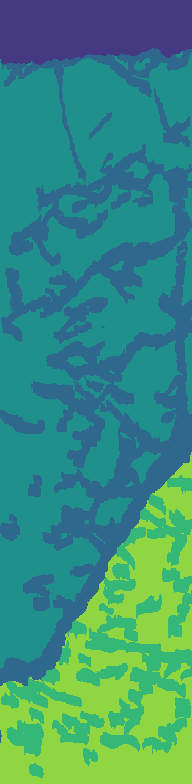}} &
\subfloat{\includegraphics[height = \ultrasoundheightA, width = \ultrasoundwidth]{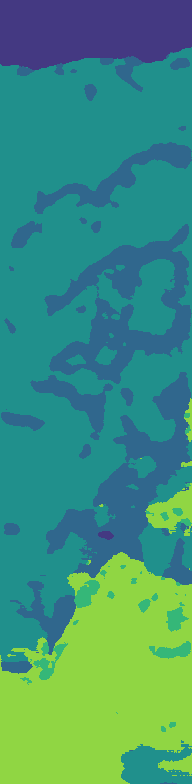}} &
\subfloat{\includegraphics[height = \ultrasoundheightA, width = \ultrasoundwidth]{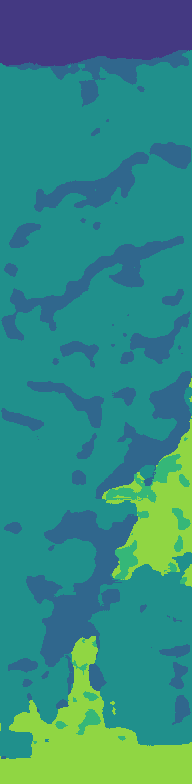}} &
\subfloat{\includegraphics[height = \ultrasoundheightA, width = \ultrasoundwidth]{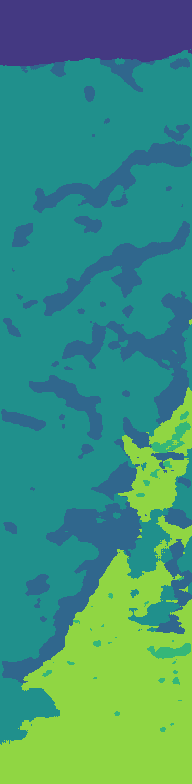}} &
\subfloat{\includegraphics[height = \ultrasoundheightA, width = \ultrasoundwidth]{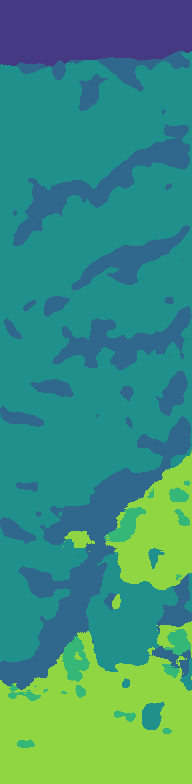}} &
\subfloat{\includegraphics[height = \ultrasoundheightA, width = \ultrasoundwidth]{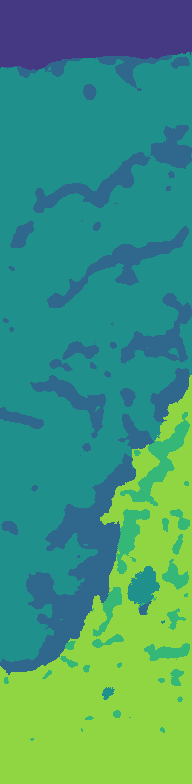}}\\
\subfloat{\rotatebox[origin=lB]{90}{ \scriptsize patient-A}} &
\subfloat{\includegraphics[height = \ultrasoundheightA, width = \ultrasoundwidth]{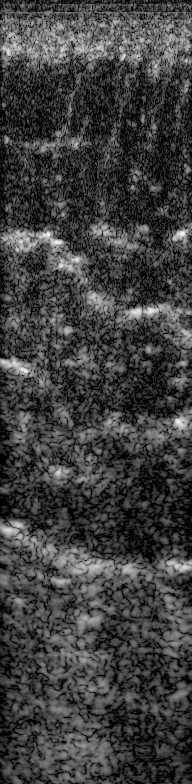}} &
\subfloat{\includegraphics[height = \ultrasoundheightA, width = \ultrasoundwidth]{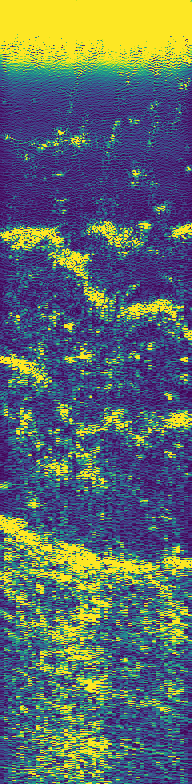}} &
\subfloat{\includegraphics[height = \ultrasoundheightA, width = \ultrasoundwidth]{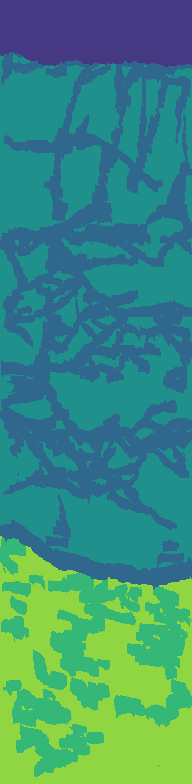}} &
\subfloat{\includegraphics[height = \ultrasoundheightA, width = \ultrasoundwidth]{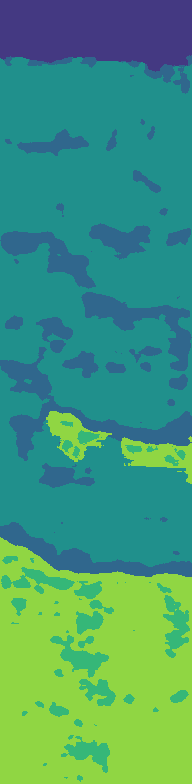}} &
\subfloat{\includegraphics[height = \ultrasoundheightA, width = \ultrasoundwidth]{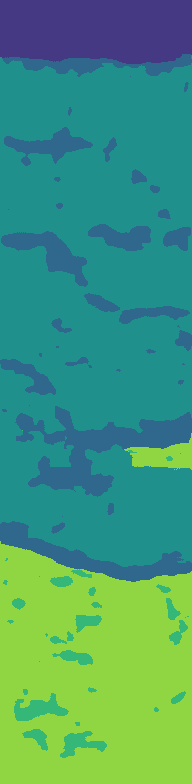}} &
\subfloat{\includegraphics[height = \ultrasoundheightA, width = \ultrasoundwidth]{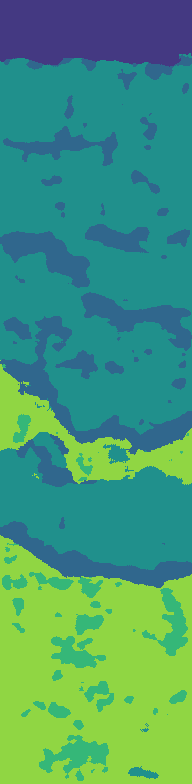}} &
\subfloat{\includegraphics[height = \ultrasoundheightA, width = \ultrasoundwidth]{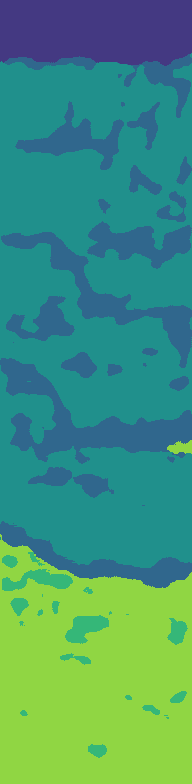}} &
\subfloat{\includegraphics[height = \ultrasoundheightA, width = \ultrasoundwidth]{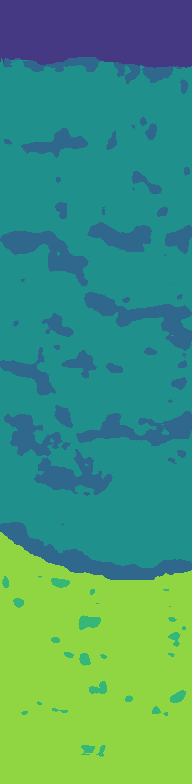}}\\
\subfloat{\rotatebox[origin=lB]{90}{ \scriptsize patient-B}} &
\subfloat{\includegraphics[height = \ultrasoundheightB, width = \ultrasoundwidth]{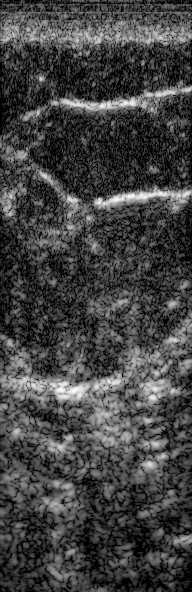}} &
\subfloat{\includegraphics[height = \ultrasoundheightB, width = \ultrasoundwidth]{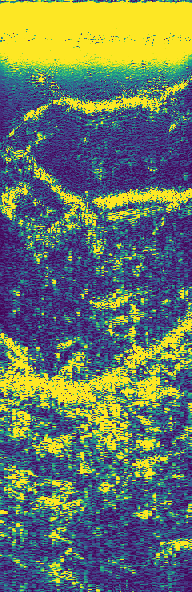}} &
\subfloat{\includegraphics[height = \ultrasoundheightB, width = \ultrasoundwidth]{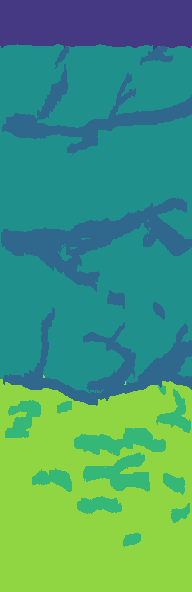}} &
\subfloat{\includegraphics[height = \ultrasoundheightB, width = \ultrasoundwidth]{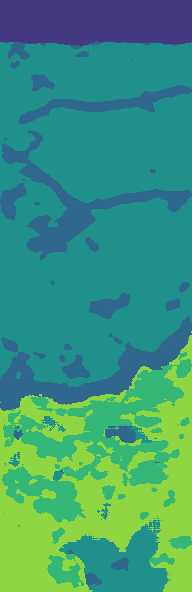}} &
\subfloat{\includegraphics[height = \ultrasoundheightB, width = \ultrasoundwidth]{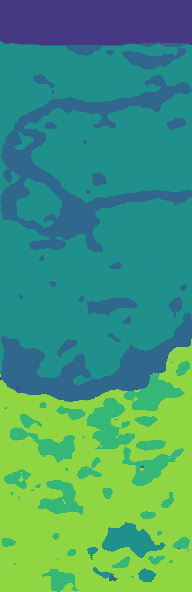}} &
\subfloat{\includegraphics[height = \ultrasoundheightB, width = \ultrasoundwidth]{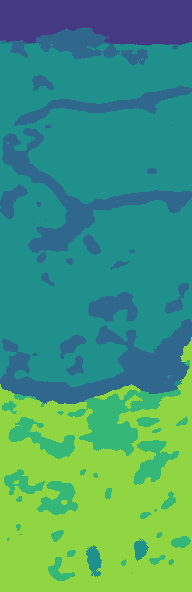}} &
\subfloat{\includegraphics[height = \ultrasoundheightB, width = \ultrasoundwidth]{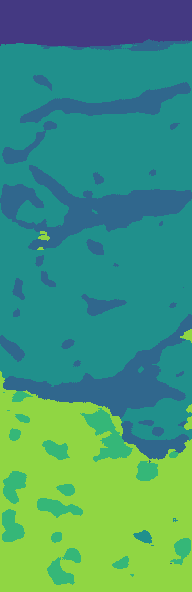}} &
\subfloat{\includegraphics[height = \ultrasoundheightB, width = \ultrasoundwidth]{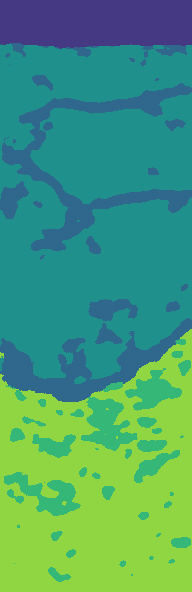}}\\
\subfloat{\rotatebox[origin=lB]{90}{ \scriptsize patient-B}} &
\subfloat{\includegraphics[height = \ultrasoundheightB, width = \ultrasoundwidth]{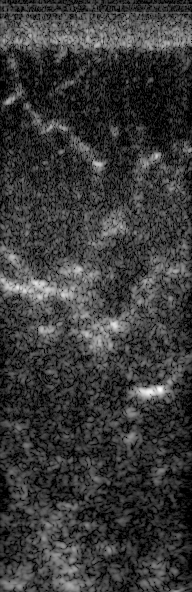}} &
\subfloat{\includegraphics[height = \ultrasoundheightB, width = \ultrasoundwidth]{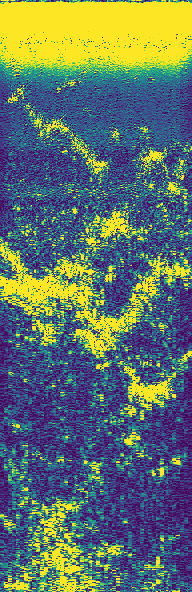}} &
\subfloat{\includegraphics[height = \ultrasoundheightB, width = \ultrasoundwidth]{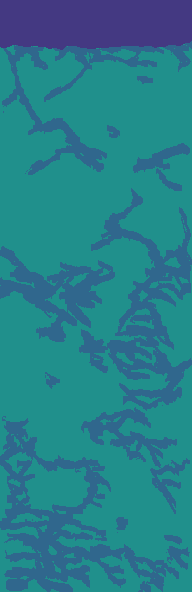}} &
\subfloat{\includegraphics[height = \ultrasoundheightB, width = \ultrasoundwidth]{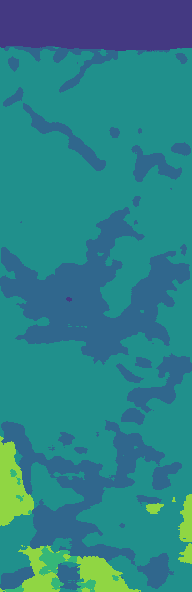}} &
\subfloat{\includegraphics[height = \ultrasoundheightB, width = \ultrasoundwidth]{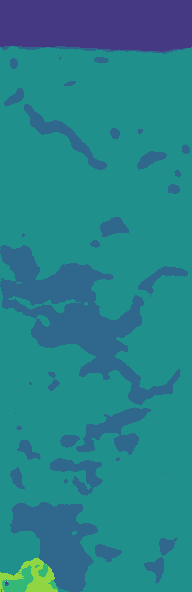}} &
\subfloat{\includegraphics[height = \ultrasoundheightB, width = \ultrasoundwidth]{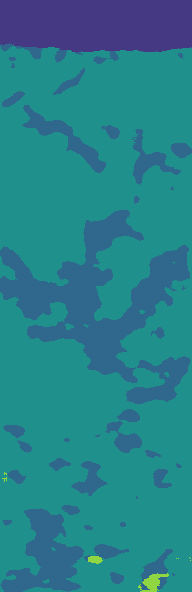}} &
\subfloat{\includegraphics[height = \ultrasoundheightB, width = \ultrasoundwidth]{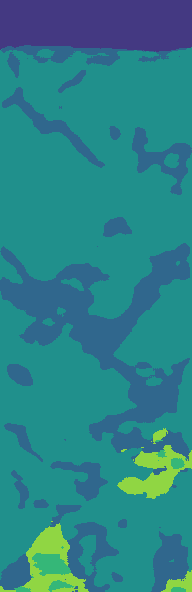}} &
\subfloat{\includegraphics[height = \ultrasoundheightB, width = \ultrasoundwidth]{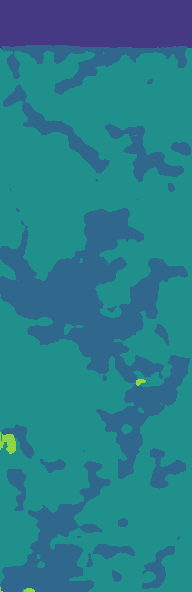}}
\end{tabular}
\caption{
\small Segmentation results on four test images with input grey and RF images ( false-color +ve/--ve heat map). Labels:
\legendsquare{RoyalPurple} skin 
\legendsquare{Emerald} fat
\legendsquare{MidnightBlue} fat fascia
\legendsquare{LimeGreen} muscle
\legendsquare{ForestGreen} muscle fascia
}
\label{fig:seg_results}
\end{figure}

\subsubsection{Results/Discussions}			
Our model performs better across all tissues classes (except skin), achieving increased mIoU accuracy (averaged across all tissue classes) by 4.5\% and 4.9\% compared to regular U-Net and AU-Net respectively. For muscle fascia, an improvement of 13\% and 16\% vs U-Net and AU-Net respectively is observed. Additionally, the U-Net and AU-Net perform better when the RF data is also provided as an input, indicating that incorporation of RF data boosts segmentation accuracy.

\begin{table}[!ht]
\centering
\caption{Segmentation Pixel-wise and mIoU scores averaged over 7 independent trials}
\label{tab:seg_results}
\begin{tabular}{|c|c|c|c|c|c|c|c|c|}
 \hline
\multicolumn{1}{|c|}{\multirow{2}{*}{CNN}} & \multirow{2}{*}{Input data} & Pixel-wise & \multicolumn{6}{|c|}{mIoU } \\\cline{4-9} 

\multicolumn{1}{|c|}{}&& \multicolumn{1}{|c|}{Acc}& mean & Skin & Fat fascia & Fat & Muscle fasica & Muscle\\
\hline
U-Net & grey &	0.746$\pm$0.011  &	0.555$\pm$0.007 &	0.923$\pm$0.002 & 0.361$\pm$0.015 &	0.699$\pm$0.017 &	0.186$\pm$0.013 & 0.605$\pm$0.016 \\
 U-Net & grey+RF&	0.755$\pm$0.008  &	0.565$\pm$0.006 &	0.926$\pm$0.002 & 0.357$\pm$0.015 &	0.706$\pm$0.010 &	0.179$\pm$0.009 & 0.657$\pm$0.015 \\
AU-Net & grey &	0.741$\pm$0.015 &	0.553$\pm$0.010 &	0.924$\pm$0.004 & 0.371$\pm$0.017 &	0.689$\pm$0.018 &	0.181$\pm$0.017 & 0.601$\pm$0.020 \\
AU-Net & grey+RF &	0.740$\pm$0.013  &	0.555$\pm$0.007 &	\bfseries{0.927$\pm$0.004} & 0.355$\pm$0.013 &	0.688$\pm$0.015 &	0.179$\pm$0.017 & 0.627$\pm$0.018 \\
W-Net & grey+RF &	\bfseries{0.769$\pm$0.007}  &	\bfseries{0.580$\pm$0.005} &	0.925$\pm$0.003 & \bfseries{0.373$\pm$0.014} &	\bfseries{0.722$\pm$0.009} &	\bfseries{0.210$\pm$0.018} &\bfseries{0.669$\pm$0.011} \\
\hline
\end{tabular}
\end{table}

\section{Experiment-2: Muscle Fascia Analysis} 
The segmentation accuracy (mIoU) for the tissue classes is visualized with respect to training epochs (see Fig. \ref{fig:Wnet_model}). We observe that most of the learning for the skin, fat, and muscle tissue classes take place in the first 20 epochs after which their values converge. The learning for the fat fascia continues until epoch 40 and for the muscle fascia the learning plateaus around epoch 55. As can be observed (refer Table \ref{tab:seg_results}), muscle fascia and fat fascia/stroma have the lowest mIoU accuracy values as compared to other tissue classes. We can infer that the muscle fascia and fat fascia tissue classes are more difficult to learn as compared to other tissues classes, with muscle fascia being the most difficult class to learn.


We conduct an experiment wherein the augmented training dataset size is gradually increased and calculate the W-Net segmentation scores on a fixed validation dataset. Results are tabulated in Table \ref{tab:dataset_size_results}, which depicts the mIoU accuracy value for the tissue classes with varying dataset sizes.






\begin{table}[!ht]
\centering
\caption{mIoU accuracy with varying augmented training dataset size.}
\label{tab:dataset_size_results}
\begin{tabular}{|c|c|c|c|c|c|c|c|}
 \hline
Augmented Training & Pixel-wise & \multicolumn{6}{|c|}{mIoU } \\\cline{3-8} 
\multicolumn{1}{|c|}{Dataset Size}& \multicolumn{1}{|c|}{Acc}& mean & Skin & Fat fascia & Fat & Muscle fasica & Muscle\\
\hline
 360 & 0.737  &	0.547 &	0.914 & 0.357 &	0.688 &	0.217 & 0.559 \\
 480 & 0.740  &	0.557 &	0.931 & 0.364 &	0.689 &	0.229 & 0.568 \\
 600 & 0.739  &	0.563 &	0.929 & 0.375 &	0.681 &	0.265 & 0.566 \\
\hline
\end{tabular}
\end{table}

\textbf{Results/Discussions}
We can see that from Table \ref{tab:dataset_size_results} adding more data improves the overall accuracy. With an increase in the augmented training dataset size from 360 to 600 train images, we observe the mIoU accuracy increases by 22\% and 5\% for muscle fascia and fat fascia/stroma respectively, while no significant increase in the mIoU accuracy for the skin, fat, and muscle tissue classes is observed. One possible conclusion would be that having more training images can help improve the segmentation accuracy of muscle fascia and fat fascia/stroma. Another possible reason for the poor segmentation performance of muscle fascia and muscle as compared to fat fascia/stroma and fat could be the loss of resolution in the ultrasound image as these tissues are anatomically deeper.


\section{Conclusion}
To the best of our knowledge, this is the first dense semantic segmentation CNN for ultrasound that attempts to label every pixel, without making use of a background label. This work is also believed to be the first application of deep learning to shallow subcutaneous ultrasound, which seeks to differentiate fascia from fat from muscle. Finally, we believe this to be the first work that uses RF (possibly any waveform) data for segmentation using deep learning techniques.

We presented a novel RF encoding branch to augment models used for ultrasound image segmentation and demonstrated the improvement in segmentation accuracy across all tissue classes by using RF data. We compared segmentation accuracy of our W-net against U-Net and AU-Net on our unique dataset. From the experiments we carried out, our W-Net achieves the best overall mIoU score and the best individual mIoU scores for muscle fascia and fat fascia/stroma (our most challenging tissue for which all methods performed the worst).

This being the first segmentation attempt in the SubQ area and considering the small dataset used for training, the results are on par with the accuracy achieved during the early stage of research in other prominent areas \cite{DBLP:journals/corr/GirshickDDM13} \cite{DBLP:journals/corr/HariharanAGM14} \cite{cnn_low_acc}. Although the results do not yet qualify for practical applications or real-world deployment of the system in its current state, we have shown that RF data can help with segmentation and warrants more research in this field.  Commercialization opportunities are presently being pursued and patents are pending.

We plan on expanding the dataset to include scans from different patients under various scan settings. We  hope to develop better data augmentation and labeling techniques. We will explore combinations with complimentary Neural Network architectures such as GANs \cite{GANs} and LSTM \cite{LSTM}, whose use in segmentation tasks have gained popularity. Finally, we will explore the utility of W-Net for other segmentation tasks involving vessels, ligaments, and needles.


\subsubsection{Acknowledgements}
We would like to thank Clarius, the portable ultrasound company, for their extended cooperation with us. We thank Dr. J. William Futrell and Dr. Ricardo Luis Rodriguez for their generous computer donation. This work used the Extreme Science and Engineering Discovery Environment (XSEDE) \textbf{\cite{xsede}} BRIDGES GPU AI compute resources at the Pittsburgh Supercomputing Center (PSC) through allocation TG-IRI200003, which is supported by National Science Foundation grant number ACI-1548562.

	



\vspace{3cm}

%
%
%
%
\bibliographystyle{splncs04}
\bibliography{main.bib}

\setlength{\ultrasoundwidth}{1.1 in}
\setlength{\ultrasoundheightA}{1.106 in}
\setlength{\ultrasoundheightB}{0.838 in}

\vspace{2cm}
\centering \textbf{\LARGE Supplementary Material} 

\begin{figure}[!ht]
\centering
\newcolumntype{C}{>{\centering\arraybackslash}m{\ultrasoundwidth}<{}}
\begin{tabular}{CCCC}
\subfloat{grey} &
\subfloat{RF} &
\subfloat{label} &
\subfloat{W-Net grey}\\
\subfloat{\includegraphics[height = \ultrasoundheightA, width = \ultrasoundwidth]{img/exp1/A1/patientA-59_gray.png}} &
\subfloat{\includegraphics[height = \ultrasoundheightA, width = \ultrasoundwidth]{img/exp1/A1/patientA-59_rf.png}} &
\subfloat{\includegraphics[height = \ultrasoundheightA, width = \ultrasoundwidth]{img/exp1/A1/patientA-59_label.png}} &
\subfloat{\includegraphics[height = \ultrasoundheightA, width = \ultrasoundwidth]{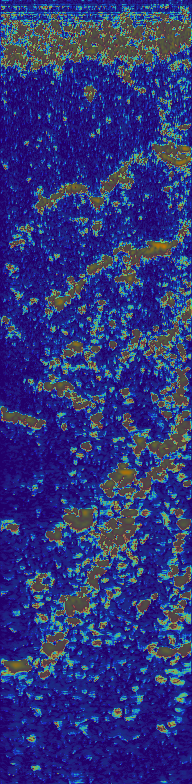}}\\
\subfloat{W-Net RF 7x3} &
\subfloat{W-Net RF 11x3} &
\subfloat{W-Net RF 21x5} &
\subfloat{W-Net RF 51x9}\\
\subfloat{\includegraphics[height = \ultrasoundheightA, width = \ultrasoundwidth]{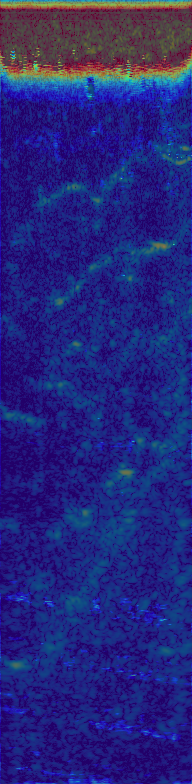}} &
\subfloat{\includegraphics[height = \ultrasoundheightA, width = \ultrasoundwidth]{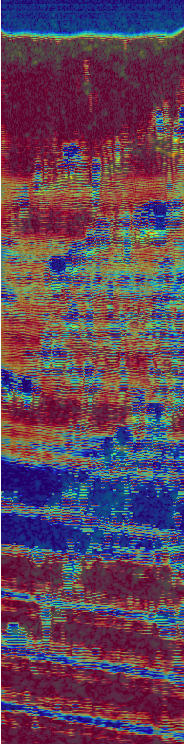}} &
\subfloat{\includegraphics[height = \ultrasoundheightA, width = \ultrasoundwidth]{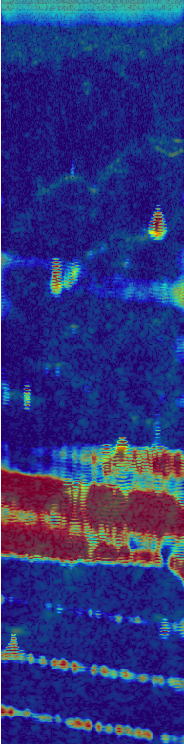}} &
\subfloat{\includegraphics[height = \ultrasoundheightA, width = \ultrasoundwidth]{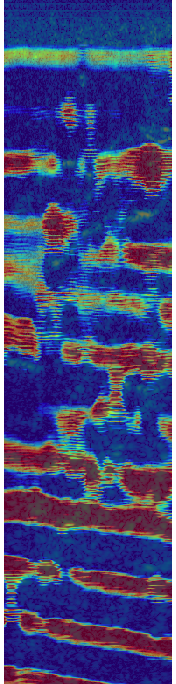}}\\
\subfloat{U-Net} &
\subfloat{U-Net with RF} &
\subfloat{AU-Net} &
\subfloat{AU-Net with RF}\\
\subfloat{\includegraphics[height = \ultrasoundheightA, width = \ultrasoundwidth]{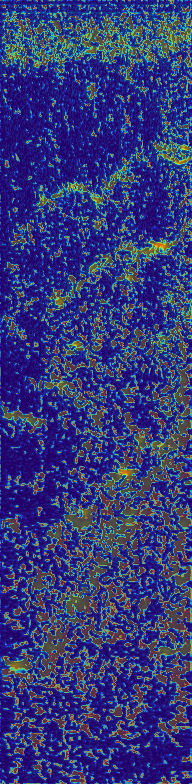}} &
\subfloat{\includegraphics[height = \ultrasoundheightA, width = \ultrasoundwidth]{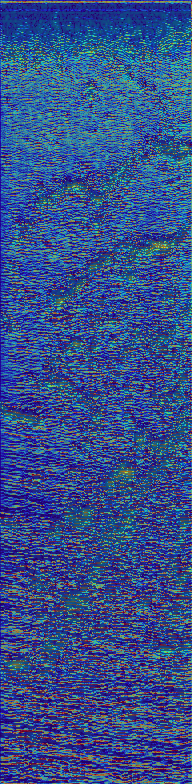}} &
\subfloat{\includegraphics[height = \ultrasoundheightA, width = \ultrasoundwidth]{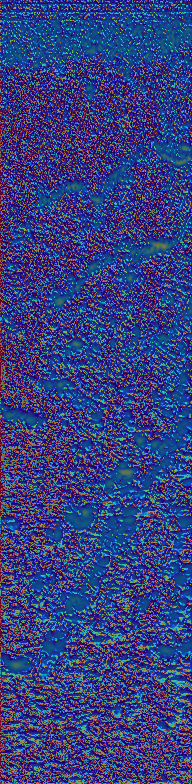}} &
\subfloat{\includegraphics[height = \ultrasoundheightA, width = \ultrasoundwidth]{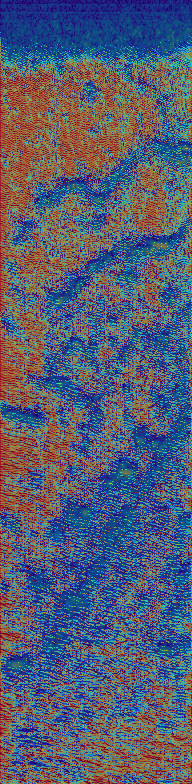}}\\
\end{tabular}
\caption{
Activation maps of the first convolutional block overlaid on top of the input grey image along with input grey, RF and label images. We can see that the W-Net grey branch activation map is noise free as compared to other CNN networks. We can observe that the W-Net's grey branch activation map is responding to skin and fascias while the RF kernel branches are responding more to particular features. Like, RF 7x3 is responding to the skin and RF 21x5 to the fat and muscle boundary region.
}
\label{fig:conv1_activation_maps}
\end{figure}

\setlength{\ultrasoundwidth}{1.200 in}
\setlength{\ultrasoundheightA}{0.91 in}
\setlength{\ultrasoundheightB}{0.838 in}

\begin{figure}[!ht]
\centering
\newcolumntype{C}{>{\centering\arraybackslash}m{\ultrasoundwidth}<{}}
\begin{tabular}{CCCCC}
&
\subfloat{grey} &
\subfloat{RF} &
\subfloat{label} &
 \\
&
\subfloat{\includegraphics[height = \ultrasoundheightA, width = \ultrasoundwidth]{img/exp1/A1/patientA-59_gray.png}} &
\subfloat{\includegraphics[height = \ultrasoundheightA, width = \ultrasoundwidth]{img/exp1/A1/patientA-59_rf.png}} &
\subfloat{\includegraphics[height = \ultrasoundheightA, width = \ultrasoundwidth]{img/exp1/A1/patientA-59_label.png}} 
& \\
\subfloat{conv1 grey} &
\subfloat{conv1 RF 7x3} &
\subfloat{conv1 RF 11x3} &
\subfloat{conv1 RF 21x5} &
\subfloat{conv1 RF 51x9}\\
\subfloat{\includegraphics[height = \ultrasoundheightA, width = \ultrasoundwidth]{img/WNet/conv1_overlayed.png}} &
\subfloat{\includegraphics[height = \ultrasoundheightA, width = \ultrasoundwidth]{img/WNet/conv1_rf_a_overlayed.png}} &
\subfloat{\includegraphics[height = \ultrasoundheightA, width = \ultrasoundwidth]{img/WNet/conv1_rf_d_overlayed.png}} &
\subfloat{\includegraphics[height = \ultrasoundheightA, width = \ultrasoundwidth]{img/WNet/conv1_rf_b_overlayed.png}} &
\subfloat{\includegraphics[height = \ultrasoundheightA, width = \ultrasoundwidth]{img/WNet/conv1_rf_c_overlayed.png}}\\
\subfloat{conv2 grey} &
\subfloat{conv2 RF 7x3} &
\subfloat{conv2 RF 11x3} &
\subfloat{conv2 RF 21x5} &
\subfloat{conv2 RF 51x9}\\
\subfloat{\includegraphics[height = \ultrasoundheightA, width = \ultrasoundwidth]{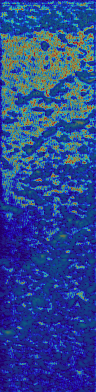}} &
\subfloat{\includegraphics[height = \ultrasoundheightA, width = \ultrasoundwidth]{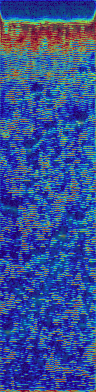}} &
\subfloat{\includegraphics[height = \ultrasoundheightA, width = \ultrasoundwidth]{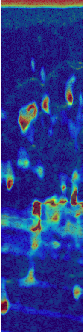}} &
\subfloat{\includegraphics[height = \ultrasoundheightA, width = \ultrasoundwidth]{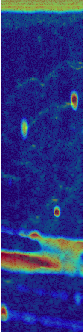}} &
\subfloat{\includegraphics[height = \ultrasoundheightA, width = \ultrasoundwidth]{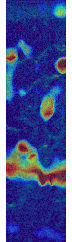}}\\
\subfloat{conv3 grey} &
\subfloat{conv3 RF 7x3} &
\subfloat{conv3 RF 11x3} &
\subfloat{conv3 RF 21x5} &
\subfloat{conv3 RF 51x9}\\
\subfloat{\includegraphics[height = \ultrasoundheightA, width = \ultrasoundwidth]{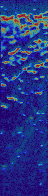}} &
\subfloat{\includegraphics[height = \ultrasoundheightA, width = \ultrasoundwidth]{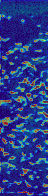}} &
\subfloat{\includegraphics[height = \ultrasoundheightA, width = \ultrasoundwidth]{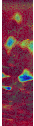}} &
\subfloat{\includegraphics[height = \ultrasoundheightA, width = \ultrasoundwidth]{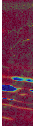}} &
\subfloat{\includegraphics[height = \ultrasoundheightA, width = \ultrasoundwidth]{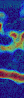}}\\
\subfloat{conv4 grey} &
\subfloat{conv4 RF 7x3} &
\subfloat{conv4 RF 11x3} &
\subfloat{conv4 RF 21x5} &
\subfloat{conv4 RF 51x9}\\
\subfloat{\includegraphics[height = \ultrasoundheightA, width = \ultrasoundwidth]{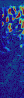}} &
\subfloat{\includegraphics[height = \ultrasoundheightA, width = \ultrasoundwidth]{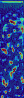}} &
\subfloat{\includegraphics[height = \ultrasoundheightA, width = \ultrasoundwidth]{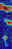}} &
\subfloat{\includegraphics[height = \ultrasoundheightA, width = \ultrasoundwidth]{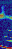}} &
\subfloat{\includegraphics[height = \ultrasoundheightA, width = \ultrasoundwidth]{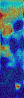}}\\
\end{tabular}
\caption{
W-Net's activation maps of the various convolutional blocks overlaid on top of the input grey image along with input grey, RF and label images. We observe that larger RF kernels are responding to more grouped regions compared to smaller RF kernels.
}
\label{fig:wnet_activation_maps}
\end{figure}

\end{document}